\documentstyle[12pt]{article}
    \date{}
\begin{document}
    
    \title{Point form relativistic quantum mechanics, antiparticles
           and exchange currents.}
    
    \author{L. Ya. Glozman}
    \maketitle
    
    \centerline{\it  The Racah Institute of Physics, The Hebrew University}
    \centerline{\it of Jerusalem, Israel and Dipartimento di Fisica}
    \centerline{\it  Nucleare e Teorica, Universita di Pavia, Italy
 \footnote{ e-mail: leonid.glozman@kfunigraz.ac.at}}
    
\begin{abstract}
It is shown that the dynamical observables calculated
with the point form relativistic quantum mechanics
incorporate effects of particle-antiparticle creation
from the vacuum by  interactions. The electromagnetic
observables obtained with the point form impulse 
approximation include contributions from the ``pair''
exchange currents that are associated with the interactions
between particles. This implies that the recently calculated
nucleon electromagnetic formfactors with the chiral constituent
quark model automatically take into account 
effects of ``pair'' exchange currents
that are associated with the Goldstone boson exchange
between the constituent quarks as well as with the confining
interaction.
\end{abstract} 

    \setcounter{page} {0}
    \vspace{1cm}

Among the various forms of relativistic quantum dynamics,
introduced by Dirac \cite{DIRAC,KP}, the point form has
some obvious advantages. It contains interactions in 4-momentum
operators, while the six generators of Lorentz transformations
(i.e. boosts and spatial rotations) are pure kinematic. This
feature is rather important and allows one to perform  manifestly
covariant calculations of different dynamical observables once
one posseses a few particle wave function in the rest frame 
\cite{KLINK1}. Very recently the latter formalism has been
applied to nucleon electromagnetic formfactors \cite{WAG}. In
that work, use has been made of nucleon wave functions obtained
within the chiral constituent quark model of ref. \cite{GPVW}
which relies on the idea of  Goldstone boson exchange between
the constituent quarks in baryons in the low-energy regime
of QCD \cite{GR,GV}. This type of interactions between the
quasiparticles (constituent quarks) allows to a simultaneous
description of the low-energy baryon mass spectrum in all
u,d,s flavor sectors (for a recent overview see ref. \cite{PANIC}).
The calculation in ref. \cite{WAG} is a straightforward one
(following all the prescription of ref. \cite{KLINK1}) and
does not involve any adjustable parameters. All the
predictions are found to be in remarkable agreement with
existing experimental data. For example, the electric proton
and neutron formfactors as well as radii fit experiment
within the error bars. Only magnetic moments and magnetic
formfactors
deviate from the experimental data by a few percent in
magnitude, though the magnetic formfactors follow the
required dipole form below 1 GeV. Given this success it is
rather important to understand a physical content of
dynamical observables calculated within the present approach.\\

Within this approach,
which is called point form relativistic impulse approximation,
 at face value only
 particle degrees of freedom are involved  
and  the
process associated with the creation of pairs from the
vacuum as well as related exchange currents are not
included \cite{KLINK1,WAG}. This impression is motivated by the fact that
the few particle (baryon) wave function, that should
be obtained in the baryon
rest frame 
and can be used to calculate any observable, 
contains only particle degrees of freedom  and
no antiparticle  degrees of freedom are  introduced.
The purpose of this letter is rather  limited, we are not going to
calculate anything, but rather aim to give a proper physical
interpretation of the results obtained in such an approach.
We demonstrate that
the dynamical observables like baryon electromagnetic formfactors,
calculated in the manifestly covariant point form approach,
do include creation of the antiparticles from the vacuum
(quark-antiquark pairs) and associated exchange two- and 
three-body currents.
In order to see it we will heavily rely on Feynman interpretation
of quantum mechanics. This is in contrast to the operator
formalism that is used in ref. \cite{KLINK1}.\\

We will begin with the brief overview of the point form
formalism. The space- and time-evolution equation in
the point form relativistic quantum mechanics is prescribed
by the covariant equation

\begin{equation}    
P^\mu \langle x|\Psi\rangle = 
\imath\hbar\partial^\mu \langle x|\Psi \rangle,
\label{SCH}
\end{equation}

\noindent
where $P^\mu$ is the 4-momentum operator and $\Psi$ is
an element of the Hilbert space. The 4-momentum operators
satisfy the standard Poincare algebra. In this approach
all four components of the 4-momentum contain interaction,
which is in contrast to e.g. the instant form of
relativistic quantum dynamics (or nonrelativistic
Schr\"odinger equation) that include interaction in the
Hamiltonian only. In the system rest frame the eq. (\ref{SCH})
becomes the Scr\"odinger type equation that describes an
evolution of the system in proper time. The spectrum of the
mass operator, $ M=\sqrt P_\mu P^\mu$, specifies the bound
as well as scattering states. The n-particle basis states
(i.e. n-particle system without interactions)
are introduced as tensor products of single-particle states
$|p_1j_1\sigma_1 \rangle \times ... \times |p_nj_n\sigma_n 
\rangle $, where
$p_i, j_i, \sigma_i$ are 4-momentum, spin and spin-projection
of the i-th particle. Under Lorentz transformation $\Lambda$ each 
 single-particle state is transformed in a standard way

\begin{equation}
U(\Lambda)|p_ij_i\sigma_i \rangle =\sum_{\sigma_i'}
D_{\sigma_i' \sigma_i}^{j_i}(R_W[p_i,\Lambda])
|\Lambda p_ij_i\sigma_i' \rangle,
\label{BOOST}
\end{equation}

\noindent
where $R_W=R_W(p,\Lambda)$ is the corresponding Wigner
rotation and $U(\Lambda)$ forms an infinite-dimensional
unitary representation of the Lorentz group. It is very
convenient to introduce the so-called velocity states
\cite{KLINK2}

\begin{equation}
|v,\vec k_i, \mu_i \rangle = 
 \sum_{\sigma_i}|p_1j_1\sigma_1 \rangle ... 
|p_nj_n\sigma_n \rangle \prod_{i=1}^n
D_{\sigma_i \mu_i}^{j_i}(R_W[k_i,B(v)]),
\label{VEL}
\end{equation}

\noindent
that ensure that under  Lorentz transformation $\Lambda$,
the system overal 4-velocity $v_\mu$ ($v_\mu v^\mu =1)$
goes to $\Lambda v$, while all internal momenta $\vec k_i$
(satisfying $\sum_i \vec k_i =0$) and all individual spins
of particles are all rotated by the same Wigner rotation.
In the expression above $B(v)$ is a boost carrying $p_i$
to $k_i=B^{-1}(v)p_i$ with $\sum_i \vec k_i =0$. These states
are very useful as they form the basis where all the individual
spins and orbital angular momenta can be coupled to the
total spin exactly as it is done in nonrelativistic
quantum mechanics.\\

The action of the various operators on velocity states
in the system that contains no interaction between
particles is specified as

\begin{equation}
M_{fr} |v,\vec k_i, \mu_i \rangle = 
  \sum_{i=1}^n \sqrt (m_i^2+{\vec k_i}^2) 
|v,\vec k_i, \mu_i \rangle
,
\label{MASS}
\end{equation}

\begin{equation}
V^\mu |v,\vec k_i, \mu_i \rangle = 
 v^\mu 
|v,\vec k_i, \mu_i \rangle
,
\label{VEL1}
\end{equation}

\noindent
where the 4-velocity operator $V^\mu$ is defined as
$P^\mu_{fr} = M_{fr} V^\mu$. The interaction between
particles in the system is introduced according to
Bakamjian-Thomas construction \cite{BT} so that this
interaction does not peturb the 4-velocity of the system
 but does perturb the 4-momentum and mass operators

\begin{equation}
P^\mu=MV^\mu,
\label{INT}
\end{equation}

\noindent
where $M=M_{fr} + M_{int}$ is the sum of the free
and interacting mass operator which should
satisfy $[V^\mu,M]=0$, $U(\Lambda) M U^{-1}(\Lambda)=M$,
in order to provide the Poincare algebra of 4-momentum
operators. In the rest frame, the 4-velocity of the system
is $v=(1,0,0,0)$ and the equation (\ref{SCH}) coincides with the
proper-time-dependent Schr\"odinger equation with the
only departure from the nonrelativistic Schr\"odiger
equation  being that the
nonrelativistic kinetic energy for each particle in the
system  should be substituted by
the relativistic one. Once the proper-time dependence
is exctracted from the equation and the wave function, one
obtains the stationary equation

\begin{equation}
(M_{fr} + M_{int}) |\Psi \rangle = {\mathcal{M}} |\Psi \rangle,
\label{EQ}
\end{equation}

\noindent
with the eigenvalue $\mathcal{M}$ and the eigenfunction $|\Psi \rangle$.
It is the latter equation that is solved variationally
 for the 3Q system in ref. \cite{GPVW}. Obviously this 
equation as well as its
solution contains only particles and there are no antiparticles.
The proper time evolution of the stationary solutions is taken
into account by the factor $\exp\left(-\imath M \tau/\hbar\right)$. 
Such
a solution combines the space - proper-time evolution of all
particles of the system in the system rest frame. It contains
only positive energy solutions that propagate forward in proper 
time.\\

In order to vizualize the physical content of  observables
obtained in the point form relativistic quantum mechanics
in the following we will rely on Feynman space-time interpretation
of quantum mechanics. Since in the rest frame the evolution
equation (\ref{SCH}) coincides with the Schr\"odinger
type equation, all the physical information, including the
bound state spectrum and the wave functions can be obtained from
the path integral  which represents a tool to
calculate the time-dependent Green function of the system.
Since in the following we consider the whole system to be
at rest  the time  which is actually used, $\tau$, is a proper 
time for the whole
interacting system, to be distinguished from the time
$t$ in the moving frame.\\

 The time-dependent Green function of the individual 
particle in the system, that describes propagation
of the particle from the space-time point $x$ to
the space time point $x'$,
can be  obtained from the integral equation 

\begin{equation}
G(x',x) = G_0(x',x) + \int d^4 x'' G_0(x',x'') V(x'') G(x'',x),
\label{TIME}
\end{equation}

\noindent
where the interaction of our particle with the other
particles of the system at the space-time point $x''$
is given by $V(x'')$ and the free Green function $G_0(x'-x)$

\begin{equation}
G_0(x'-x) = -\imath \int \frac{d{\vec k_i}}{(2\pi)^3 2E_k}
\exp \left (\imath {\vec k_i}({\vec x'} - {\vec x})\right )
\exp \left (-\imath E_k (\tau' - \tau)\right )
\Theta (\tau' - \tau),
\label{FREE}
\end{equation}

\noindent
where
$ E_k=\sqrt{{\vec k_i}^2 + m_i^2}$.\\

From now on we shall follow a very elegant lecture
`` The reason for antiparticles'' , given
by Feynman \cite{F} which is devoted to Dirac.
He proves there the following: ``If we insist that
particles can only have positive energies, then
you cannot avoid propagation outside the light
cone. If we look at such propagation from a
different frame, the particle is traveling backwards
in time: it is an antiparticle. One man's virtual
particle is another man's virtual antiparticle.''\\

Consider for simplicity the amplitude $G(x',x)$ that
is quadratic in interaction $V(x)$. Such an amplitude
contains a free propagation of the particle to the
point $x_1$, interaction $V(x_1)$ with the other particles
of the system at this point, then a free propagation
of the particle from this point to the point $x_2$,
interaction $V(x_2)$ at that point and a free propagation
afterwards. This amplitude involves an integration over
all possible free trajectories that connect points
$x_1$ and $x_2$, that is an integration over 3-momentum
$\vec k_i$ of the {\it virtual} particle. Then we have to
integrate over all possible $ x_1$ and $ x_2$ provided
that $\tau_2 > \tau_1$:

\begin{equation}
\int d^4 x_1 d^4 x_2 \Theta (\tau_2 - \tau_1)
 \int \frac{d{\vec k_i}}{(2\pi)^3 2E_k}
\exp \left (\imath {\vec k_i}({\vec x_2} - {\vec x_1})\right )
\exp \left (-\imath E_k (\tau_2 - \tau_1)\right ) 
a(x_1) b^*(x_2),
\label{FEYN}
\end{equation}
 
\noindent
where  the
functions $a(x_1)$ and $b^*(x_2)$ contain all
the information about the interaction at the
points $x_1$ and $x_2$ respectively,

\begin{equation}
a(x_1) = V(x_1) \phi_0 (x_1) \sqrt{2E_K},
\end{equation}

\begin{equation}
b(x_2) = V(x_2) \phi_0 (x_2) \sqrt{2E_K},
\end{equation}

\noindent
and $\phi_0(x_1)$ and $\phi_0(x_2)$ are initial and final
plane waves.\\

The integration over $\vec k_i$ can be turned into the
integration over $\omega=E_k$ and defining $F(\omega)=0$ for any 
$\omega < m_i$, the integral over $\vec k_i$ in (\ref{FEYN})
is reduced to the following integral:

\begin{equation}
f(\tau)=\int_0^\infty e^{-\imath \omega \tau} F(\omega) d\omega,
\label{OMEGA}
\end{equation} 

\noindent
where the dependence on $x_1$ and $x_2$ is absorbed into
$F(\omega)$.  Feynman makes  use of the following mathematical
theorem. If the function $f$ can be Fourier decomposed into the
{\it positive frequencies only}, like in the equation above, then
$f$ cannot be zero for any finite range of $\tau = \tau_2 - \tau_1$,
unless trivially it is zero everywhere. An immediate implication
is that at the given $x_1$ the integral over $\vec k_i$
in (\ref{FEYN}) cannot be zero for $x_2$ that is outside the
light cone of $x_1$. Hence the integral (\ref{FEYN}) necessarily
involves the {\it nonzero} amplitudes that contain the space-like
interval between $x_2$ and $x_1$. If the interval is space-like,
then the time order of the events is frame dependent. 
While in
the rest frame one always has $\tau_2 > \tau_1$, i.e. 
all the virtual amplitudes describe propagation of particles 
forward
in time, some of these  amplitudes are
seen in the moving reference frame as virtual particle
 that propagates
backward in time, i.e. as antiparticle moving forward in time. 
This is a pure
kinematics and cannot be circumvented.\\

 It is important
to realize, however, that this is a result of {\it interactions} 
of our
particle with the other ones. Without interaction, i.e.
when a particle propagates freely from the very beginning to the
very end, all the intervals are time-like, i.e. the time order of
the events is frame-independent. The free particle is seen
as a particle in all frames.  That is the Lorentz transformation 
of the system of particles without interaction (\ref{BOOST}).
However once we boost the wave function of the system
 {\it with}
interactions, where the individual 4-momenta of particles
{\it are not defined} (such a wave function is a very
complicated superposition of the wavefunctions of the type
(\ref{VEL})) and which necessarily includes the space-like
intervals for every {\it virtual}  particle, these kinematical boosts 
  include transformations of what we consider
as virtual particles in the rest frame, to what which should be
considered as {\it virtual}
antiparticles in  moving reference frame, i.e. it transforms
the virtual amplitudes for positive energy particles
with $\tau_2 > \tau_1$ to amplitudes with
$t_2 < t_1$ in the moving frame. Once it is done, the
wave function in the moving frame contains an admixture
of {\it virtual} antiparticles, that propagate forward in time,
though it is not seen explicitly in the time independent operator
formalism \cite{KLINK1}.
Schematically it is shown in Fig. 1. What we observe as
interactions between particles in the rest frame, Fig. 1a, 
corresponds to creation from the vacuum (and annihilation) of the
antiparticles by the {\it interaction} in the moving frame, 
Fig. 1b.
The magnitude of those virtual amlitudes (with the space-like
intervals) is completely specified by the interaction
between particles.
\\

The argument does not change if one assumes that at the point
$x_1$ the particle of the system is coupled to the
external field (e.g. electromagnetic one), propagates freely
to the point $x_2$ and interacts there with other particles.
Since the matrix element of the electromagnetic current,
that is manifestly covariant  \cite{KLINK1},
involves the system that  either in initial, or in final states
(or in both initial and final states) is not at rest, it
automatically includes the ``pair'' currents, depicted in Fig. 2,
in addition to the standard 1-body currents that contain no
virtual antiparticles.
Specifically, keeping in mind the model of ref. \cite{GPVW},
this implies that the numerical result of ref. \cite{WAG} does 
include the ``pair'' exchange currents associated with the confining
interaction between quarks as well as with the Goldstone boson
exchange.\\

We have used for simplicity only the amplitude that
is quadratic in the interaction. It is obvious, however,
that the same effect is present in all higher order
amplitudes that contribute in the nonperturbative
calculation.\\

That the ``pair'' exchange currents are very important
is known from nuclear physics for a long time \cite{RIS,MAT}.
In that case, i.e. when the nonrelativistic wave function
is used and no Lorentz boosts are applied to wave function,
these ``pair'' currents are introduced explicitly using
the nonrelativistic $v/c$ expansion of the corresponding
Feynman diagram \cite{CR} so that the effect of these
``pair'' currents is absorbed into two- and three-body (etc.)
operators that act in the Hilbert space of nonrelativistic
wave function. These two- and many-body operatorts are
used in addition to the nonrelativistic one-body
current operator. The nonrelativistic one-body current represents
the nonrelativistic impulse approximation. In the
nonrelativistic impulse approximation only one particle
(that is struck by the external field) is involved and
other particles of the system are spectators. This
is  specified by the momentum delta-functions for
the spectator particles that guarantee for this particles
equality of their initial and final momenta
$\vec k_i' = \vec k_i$ {\it both in the rest and in
any moving frame} as well as by the fact that the spins
of these spectator particles {\it are not affected}. The 
momentum
transfer from the struck particle to the spectator one
is provided only via wave function of the system that
contains only particles and is
exactly the same in the rest and moving frames. 
The two- and many-body ``pair'' current operators 
affect both the spins and momenta
of two (many) particles and the momentum transfer from
one particle to another one is provided by the operator,
but not by the nonrelativistic wave function.\\

The point form relativistic impulse approximation 
\cite{KLINK1} involves one struck
particle and all other ones to be spectators only 
in that reference
frame (e.g. the Breit one) where the matrix element is
calculated. The wave function of the system
in the initial and final states do not coincide with
that one in the rest frame. The former wave functions 
necessarily contain effects of creation of pairs from the vacuum
by  interaction between particles.
These effects  are implicitly introduced
by boosting the wave function from the system rest
frame, though boosts by itself are pure
kinematic. It is important that this effect is completely
specified by the interaction between particles
that is used in the given model for the interaction
in the system rest frame.
 The boosting affects spins of the ``spectator''
particles (which is provided by the corresponding Wigner
rotations) as well as their momentum distributions.
The latter is explicitly seen from the fact that instead
of the condition $\vec k_i' = \vec k_i$, which is valid
in the nonrelativistic impulse approximation, one actually
has $\vec k_i' = B^{-1}(v_{final})B(v_{initial})  \vec k_i$.
Stated differently, boosting of the wave function from the system
rest frame to the moving one, provides a set of two- and
many-body operators that act on the  space of the {\it rest
frame wave functions}.
These {\it include} effects of ``pair'' 
exchange currents introduced in the context of nonrelativistic
nuclear physics. Contrary to the nonrelativistic scheme,
however, the $v/c$ expansion is not used and the ``pair''
currents are taken into account to all orders in interaction.
Returning then to the calculation of nucleon electromagnetic
formfactors in ref. \cite{WAG} one concludes that the ``pair''
exchange currents that are associated with the Goldstone
boson exchange between the constituent quarks and confining
interaction are automatically included.\\

I am greateful to S. Boffi, W. H. Klink and D.O. Riska for 
a quick response and
to A. Gal and the Nuclear Theory Group of
the Racah Institute for hospitality and support during my 
visit to Jerusalem where the present work has been done.

\bigskip
\bigskip
\center{FIGURE CAPTIONS}

\bigskip
\noindent
Fig. 1 (a) Time evolution of  particles as seen in the system
           rest frame; (b) the same as seen in the moving frame.
           The dashed line represent interactions.

\bigskip
\noindent
Fig. 2  ``Pair'' current contributions to the electromagnetic
        observables, that involve creation of particle-antiparticle pairs
       by the interaction between particles in the system and by the
      electromagnetic field.     
    
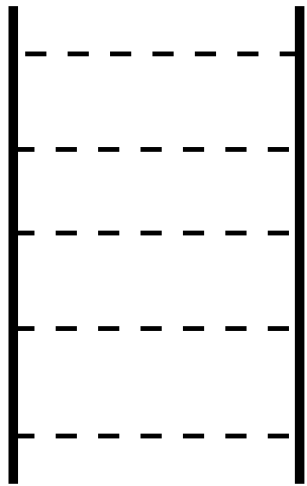
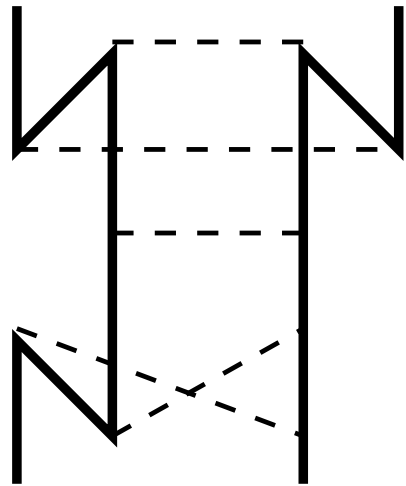
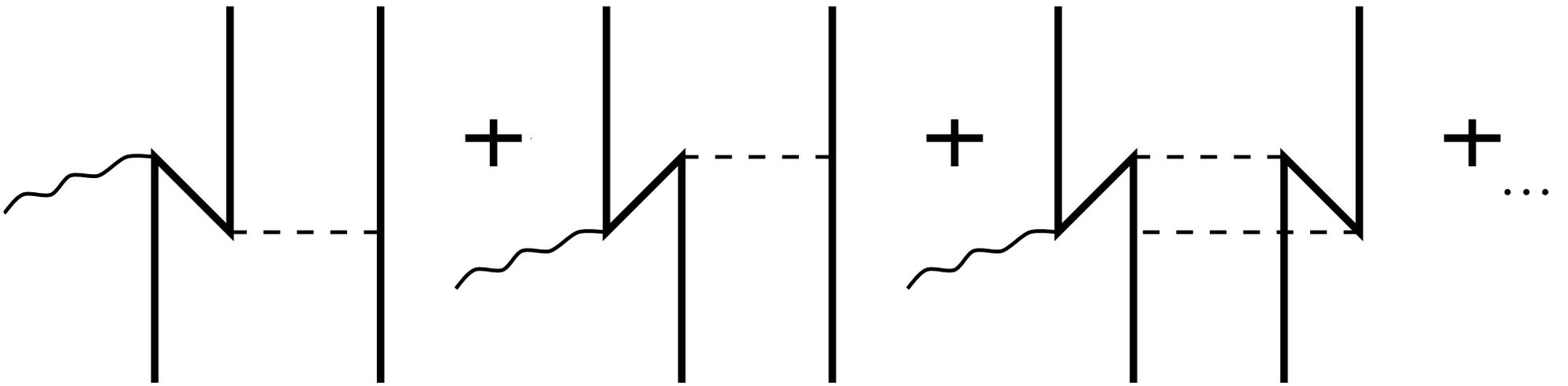
\end{document}